\documentclass[%
preprintnumbers,
nofootinbib,
 amsmath,amssymb,
 aps,
 showkeys,
]{revtex4-2}

\bibliography{report}

\usepackage{graphicx}
\usepackage{dcolumn}
\usepackage{bm}

\begin{document}

\preprint{KEK-TH-2630, J-PARC-TH-0306}

\title{On the time-dependent Aharonov-Bohm effect \\
and the 4-dimensional Stokes theorem}

\author{Masashi Wakamatsu}
 \email{wakamatu@post.kek.jp}
\affiliation{%
KEK Theory Center, Institute of Particle and Nuclear Studies,
High Energy Accelerator Research Organization (KEK),
Oho 1-1, Tsukuba, 305-0801, Ibaraki, Japan
}%




\date{\today}

\begin{abstract}
The time-dependent Aharonov-Bohm (AB) effect considers the situation
in which the magnetic flux inside the solenoid changes time-dependently.
Different from the standard AB-effect, the problem is unexpectedly subtle
and not easy to solve without any doubt, which is the reason why 
it is still in a state of unsettlement even theoretically. 
The difficulty originates from the fact that its theoretical analysis requires
line-integrals of the time-dependent vector potential along  paths in the 
4-dimensional Minkowski space. Owing to the 4-dimensional Stokes theorem, 
this closed line-integral of the vector potential can be related to the integral of 
the electric and magnetic fields over the 2-dimensional area, the boundary 
of which is given by the above-mentioned closed path.     
The central controversy concerns the success or failure of the claim 
by Singleton and collaborators based on the 4-dimensional Stokes theorem,
which states that the time-dependent part of  the AB-phase shift due to 
the magnetic vector potential is precisely cancelled by the effect of induced 
electric field generated by the time-variation of the magnetic flux.
In the present paper, we carefully reanalyze their cancellation argument by going 
back to the basic quantum mechanical analysis of the charged particle motion 
under the presence of the time-dependent external electromagnetic potential
combined with more careful treatment of the 4-dimensional Stokes theorem. 
Careful analysis of the 4-dimensional Stokes theorem shows that
the cancellation argument by Singleton et al. is partially 
correct but their central claim that only the time-independent part of the magnetic
field contributes to the AB-phase shift so that there is no time-dependent AB-effect 
is not justified, thereby supporting likely existence of the time-dependent AB-effect. 
\end{abstract}

\keywords{Time-dependent Aharonov-Bohm effect, 
4-dimensional Stokes theorem, quantum mechanics, gauge transformation}


\pacs{01.55.+b, 03.65.-w, 03.65.Vf,11.15.-q}

\maketitle


\section{\label{sec1}Introduction}

The time-dependent Aharonov-Bohm effect considers what shift of phase one 
would observe if one varies the magnetic flux inside a solenoid time-dependently.
So far, no clear experimental evidence of the time-dependent AB-phase shift
has been found.  What's worse, its existence or non-existence  has been 
a subject of unresolved theoretical controversy. 
The central controversy concerns the validity or failure of the claim by Singleton 
and his collaborators, who insist that the time-dependent part of the AB-phase shift due 
to the magnetic vector potential is precisely cancelled by the phase change generated by 
the induced electric field due to time-variation of the magnetic field  inside
the solenoid and that only the AB-phase shift corresponding to the time-independent
part of the magnetic field remains \cite{SV2013,MS2014,AS2016}. 
On the other hand, there exist not a few literature \cite{LYGC1992,ADC2000,
GNS2011,JZWLD2017,CM2019}, in which they insist 
that the time-dependent AB-effect {\it does} exist and it can in principle be observed.
Unfortunately, it is difficult to say that the authors of these papers convincingly
succeeded to show exactly where and why the cancellation argument by Singleton et al.
is not justified, while they insists the existence of the time-dependent AB-effect.
The purpose of the present paper is to fill this gap.
As faithful as possible to the basic principle of quantum mechanics, i.e. the
Schr\"{o}dinger equation for a charged particle motion under the presence of 
magnetic flux of a solenoid combined with more careful analysis of the 4-dimensional
Stokes theorem, we try to clarify the reason why Singleton et al.'s cancellation 
argument would not  be justified at least in the form as claimed in their papers.

\vspace{2mm}
The paper is organized as follows.
First, in sect.\ref{sec2}, we briefly review the cancellation argument by Singleton and
collaborators \cite{SV2013,MS2014,AS2016}. 
In particular, we remind readers of the two covariant expressions of
the time-dependent AB-phase shift discussed in their paper.
The 1st expression is given in the form of the closed line integral of the
4-vector potential in the 4-dimensional Minkowski space, whereas the 2nd 
expression is given in the form of the surface integral of the electromagnetic
field strength tensor in the 4-dimensional Minkowski space. 
The equivalence of the two expressions is taken for granted on the basis of 
the 4-dimensional Stokes theorem.  
In view of the fact that Singleton et al's cancellation argument is primarily based on 
the 2nd covariant expression, we shall proceed in the following way.
First, in sect.\ref{sec3}, by going back to the basic equation of quantum mechanics,
i.e. the Schr\"{o}dinger equation describing the motion of a charged particle
in the presence of the infinitely-long solenoid with the time-dependent
magnetic flux, we try to provide a firm theoretical basis for the 1st covariant 
expression for the time-dependent AB-phase shift. 
This confirms that the AB-phase shift is proportional to the closed space-time
line integral even in the case where the magnetic flux is changing time-dependently.
Next, in sect.\ref{sec4}, we carefully check whether the use of the 2nd forms of
covariant expression really predicts the cancellation between the time-dependent
part of the AB-phase shift. We show that the careful treatment of
the surface integral in this expression partially confirms the cancellation argument
by Singleton et al., but nevertheless it does not support the validity of their 
central conclusion.
In sect.\ref{sec5}, we argue that the fully satisfactory theoretical prediction
of the time-dependent AB-phase shift is a hard theoretical challenge 
contrary to naive expectation, but anyhow we try to give a reasonable approximate 
answer to the problem under some suitably prepared experimental setting.
Finally, in sect.\ref{sec6}, we briefly summarize our findings in the present paper.

\section{\label{sec2}Brief review of Singleton et al.'s cancellation argument}

The argument by Singleton et al. starts with the observation that, in the
familiar setting of the {\it time-independent} magnetic flux, the following two forms
of covariant expression for the AB-phase shift are equivalent.  (This is unquestionably
correct, at least in the situation where the 4-vector potential $A^\mu$ is 
time-independent, and no electric field is present in the system.)
\begin{eqnarray}
 &\,& \mbox{\tt (I)} \ \mbox{\tt 1st form : }  
 \phi_{AB} = - \,e \,\oint \,A_\mu \,d x^\mu 
 =  e \,\left[ \,- \,\int_{t_1}^{t_2} \,A^0 \,d t
 +  \oint \,\bm{A} \cdot d \bm{x} \,\right] , \label{Eq:1st_form} \hspace{6mm} \\
 &\,& \mbox{\tt (II)} \  \mbox{\tt 2nd form : } 
 \phi_{AB}  =  - \,
 \frac{e}{2} \,\int \,F_{\mu \nu} \,
 d x^\mu \wedge d x^\nu
 =  e \,\left[ \,\int \,\bm{E} \cdot d \bm{x} \,d t
 + \int \,\bm{B} \cdot d \bm{S} \,\right]. \hspace{8mm} 
 \label{Eq:2nd_form}.
\end{eqnarray}
(In the present paper, we use the natural unit $\hbar = c = 1$, and the mass
and the charge of the electron is denoted as $m$ and $- \,e$ with $e > 0$.)
Here, $A^\mu = (A^0, \bm{A})$ is a 4-vector potential, whereas 
$F_{\mu \nu} \equiv \partial_\mu \,A_\nu - \partial_\nu \,A_\mu$ is the
familiar electromagnetic field strength tensor. 
The electric field $\bm{E}$ and magnetic field $\bm{B}$ are expressed
as $E^i = F^{i 0}$ and $B^i = - \,\frac{1}{2} \,\epsilon^{i j k} \,F^{i j}$
in terms of the field strength tensor.
Since the 4-dimensional Stokes theorem represented as
\begin{equation}
 \oint \,A_\mu \,d x^\mu \ = \ \frac{1}{2} \,\int \,F_{\mu \nu} \,
 d x^\mu \wedge d x^\nu  . \label{Eq:4-dim_Stokes_theorem}
\end{equation}
is believed to be a very general mathematical identity, Singleton et al. assume
that it holds also in the case of time-varying magnetic flux.
Then, based on the form (II) of the covariant expression,  
they argued that the {\it time-dependent part} of the AB-phase 
due to the magnetic
flux is precisely cancelled by the effect of induced electric field \cite{SV2013,MS2014}.

\vspace{3mm}
In the following, we briefly review the outline of their demonstration.
Consider an infinitely-long solenoid of radius $R$ directed to the $z$-direction,
through which uniform but time-dependent magnetic flux is penetrating.  
The corresponding magnetic field distribution is expressed as
\begin{eqnarray}
 \bm{B} (\bm{x}, t) \ = \ \left\{
 \begin{array}{cc}
 B (t) \,\bm{e}_z \ & \ \ \ (\rho < R) , \\
 0 \ & \ \ \ (\rho \geq R) .\\
 \end{array} \right. 
\end{eqnarray}
Here, the position vector $\bm{x}$ is expressed as $\bm{x} = (\rho, \phi, z)$ in the 
cylindrical coordinate with $\rho = \sqrt{x^2 + y^2}$.
The vector potential reproducing the above magnetic field distribution is given by
\begin{eqnarray}
 \bm{A} (\bm{x}, t) \ = \ \left\{ \begin{array}{cc}
 \frac{\rho \,B(t)}{2} \,\,\,\bm{e}_\phi
 \ & \ \ \ (\rho < R) , \\
 \frac{R^2 \,B (t)}{2 \,\rho} \,\,\bm{e}_\phi \ & \ \ \ 
 (\rho \geq R)  . \\
 \end{array} \right. \hspace{6mm}
\end{eqnarray}
Note that one can set $A^0 = 0$ in the present setting of the problem.
(See the discussion in the next section.)
Then, from $\bm{E} = - \,\nabla A^0 - \frac{\partial \bm{A}}{\partial t} = - \,
\frac{\partial \bm{A}}{\partial t}$, it follows that 
\begin{eqnarray}
 \bm{E} (\bm{x}, t)  =  - \,\frac{\partial}{\partial t} 
 \bm{A} (\bm{x}, t)  =  \left\{ \begin{array}{cc}
 - \,\frac{\rho \dot{B} (t)}{2}
 \,\bm{e}_\phi \ & \  (\rho < R) , \\
 - \,\frac{R^2 \dot{B} (t)}{2 \,\rho} 
 \,\bm{e}_\phi \ & \  (\rho \geq R) , \\
 \end{array} \right.  \label{Eq:E-field}
\end{eqnarray}
with $\dot{B} (t) \equiv \frac{d B (t)}{d t}$. This $\bm{E} (\bm{x}, t)$ 
is nothing but the {\it induced electric field} due to the time-varying magnetic field.
Then, for a closed space-time path encircling the solenoid, they get
for the AB-phase shift
\begin{eqnarray}
 \phi_{AB} \!&=&\! \frac{e}{\hbar} \,\left[ \,
 \int \bm{E} \cdot d \bm{x} \,d t \ + \ 
 \int \bm{B} \cdot \,d \bm{S} \,\right] \nonumber \\
 \!&=&\! \frac{e}{\hbar} \,\int \,\left(
 - \,\frac{\partial \bm{A}}{\partial t} \right) \cdot d \bm{x} \,d t
 \ + \ \frac{e}{\hbar} \,\oint \bm{A} \cdot d \bm{x} 
 \ = \  - \,\frac{e}{\hbar} \,\oint \,\bm{A} \cdot d \bm{x} \ + \ 
 \frac{e}{\hbar} \,\oint \,\bm{A} \cdot d \bm{x}
 \ = \ 0 . \hspace{5mm}
\end{eqnarray}
It was interpreted to show that the effect of induced electric field precisely
cancels the effect of magnetic vector potential.

\vspace{3mm}
Their argument goes further as follows.  
From the sourceless Maxwell equation
\begin{eqnarray}
 \nabla \times \bm{B} \ = \ \frac{\partial}{\partial t} \,\bm{E} ,
\end{eqnarray}
which holds in the domain $\rho \neq R$, i.e. in the region excluding the surface of 
the solenoid, combined with the relation $\nabla \times \bm{B} = 0$ in the same region, 
it follows the constraint
\begin{eqnarray}
 \dot{\bm{E}} \ \equiv \ 
 \frac{\partial}{\partial t} \,\bm{E} \ = \ 0 .
\end{eqnarray}
Using the explicit form of $\bm{E} (\bm{x}, t)$ given by 
Eq.(\ref{Eq:E-field}), this gives
\begin{eqnarray}
 0  \ = \ \dot{\bm{E}}
 \ = \ 
 \left\{ \begin{array}{cc}
 - \,\frac{\rho \,\ddot{B} (t)}{2} 
 \,\,\bm{e}_\phi \ & \ \ \ 
 (\rho < R) , \\
 - \,\frac{R^2 \,\ddot{B} (t)}{2 \,\rho} 
 \,\,\bm{e}_\phi \ & \ \ \ 
 (\rho > R) , \\
 \end{array} \right.  \hspace{6mm}
\end{eqnarray}
which in turn dictates that $\ddot{B} (t) \equiv d^2 B(t) / d t^2 = 0$ so that $B (t)$ must 
be a linear function in time,
\begin{eqnarray}
 B (t) \ = \ B_0 \ + \ B_1 \,t .
\end{eqnarray}
(They stated that more general time-dependence can be considered,
for example, by introducing non-uniform magnetic field 
distribution inside the solenoid \cite{MS2014}.)
The above form of $B (t)$ dictates that the vector potential generating 
the above $B (t)$ also consists of the two part as
\begin{equation}
 \bm{A} (\bm{x}, t) = \bm{A}_0 (\bm{x})  \ + \ \bm{A}_1 (\bm{x}, t) ,
\end{equation}
and it gives
\begin{equation}
 \bm{B} (\bm{x}, t) = \nabla \times \bm{A} (\bm{x}, t) = 
 \bm{B}_0 (\bm{x}) \ + \ \bm{B}_1 (\bm{x}, t)
 \end{equation}
with
\begin{equation} 
 \bm{B}_0 (\bm{x}) \ = \ B_0 \,\bm{e}_z, \ \ \ 
 \bm{B}_1 (\bm{x}, t)  \ = \ 
 B_1 \,t \,\,\bm{e}_z .
\end{equation}
Since $\bm{E} = - \,\partial \bm{A} / \partial \,t$, only $\bm{A}_1$ part induces
electric field. Hence, the time-dependent part of the magnetic field, 
i.e. $\bm{B}_1 (\bm{x}, t)$ part,  to the AB-phase shift, is cancelled by the effect of
the induced electric field, so that only the time-independent AB-phase shift by
the $\bm{A}_0$ part remains. This is the central conclusion by Singleton et al.

\vspace{4mm}
\begin{figure}[ht]
\begin{center}
\includegraphics[width=7.0cm]{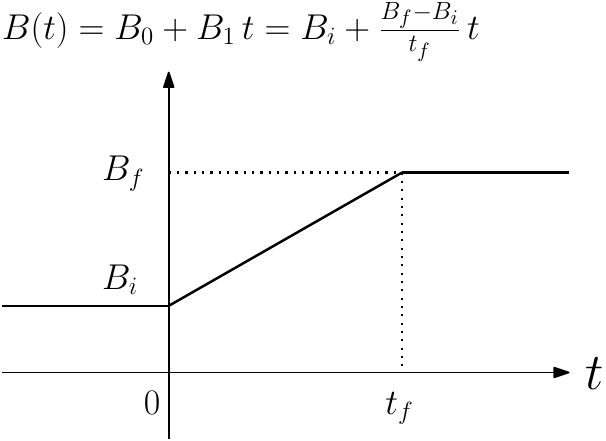}
\caption{A simple example of time-dependent magnetic flux.}
\label{Fig:time-dep.B}
\end{center}
\end{figure}

We feel that there is a little odd about the above observation.
To understand why, let us consider the setting in which the time-dependence of $B (t)$
is given as shown in the Fig.\ref{Fig:time-dep.B}.
That is, suppose that, before the initial time $t = t_i = 0$,  the magnetic field inside 
the solenoid is kept a constant value given by $B_i$.  After $t = t_i$, the magnetic
field is increased linearly with time until it reaches the value $B_f$ at the time $t = t_f$.
After the time $t = t_f$, the magnetic field is kept a constant value $B_f$. 
If Singleton et al.'s cancellation argument were correct, one might predict for
the AB-phase shift
\begin{eqnarray}
 \phi_{AB} \,(t < 0) &=& e \,\pi R^2 \,B_i
 \ = \ e \,\Phi_i ,\\
 \phi_{AB} \,(0 < t < t_f) &=& e \,
 \Phi_i \ \ \bf{?}  \\
 \phi_{AB} \,(t > t_f) &=& e \,\Phi_f \ \ \bf{?}  \ \ .
\end{eqnarray}
That is, if one measures the AB-phase shift before $t < t_i$, it is undoubtedly given 
by $e \,\Phi_i = e \,\pi R^2 \,B_i$. What AB-phase shift does one observe
between $t_i$ and $t_f$ ? If Singleton et al.'s argument is correct,
the answer would be given by $e \,\Phi_i$ since the effect of the $B_1$ part of $B (t)$
is precisely cancelled by the effect of the induced electric field.
A natural next question is what AB-phase shift does one observe after the time $t_f$ ?
Intuitively, the answer should be given by $e \,\Phi_f$, since the problem reduces to the
standard time-independent AB-effect with the constant magnetic 
flux $\Phi_f = \pi R^2 \,B_f$.
However, if this is the case, there must be a sudden jump of the observed AB-phase
shift from $e \,\Phi_i$ to $e \,\Phi_f$ at the time $t = t_f$, which we think is unlikely.
It appears to us that this simple thought experiment already indicates
internal contradictions of their cancellation argument.

\vspace{3mm}
If Singleton et al.'s prediction were not justified, what would be the reason ?
Some time ago, this question was addressed in the paper by Jing et al. \cite{JZWLD2017}.
They claim that Singleton et al's conclusion is not justified and argue in somewhat ad hoc 
manner that the time-dependent AB-effect can be 
explained by the effective Hamiltonian of the following form,
\begin{equation}
 H \ = \ \frac{1}{2 \,m} \,\left[ \,\bm{p} \ + \ e \,\bm{A} (\bm{x}, t) \,\right]^2
  \ + \ e \,\int^{\bm{x}}_{\bm{x}_0} \,\frac{\partial \bm{A} (\bm{x}^\prime,t)}{\partial t}
  \cdot d \bm{x}^\prime,
\end{equation}
which contains in it a path-dependent line-integral term. 
Based on this effective Hamiltonian, they conclude that the AB-phase shift
corresponding to time-dependent magnetic flux $\Phi (t)$ is given by
\begin{equation}
 \phi_{AB} \ = \ \oint  \bm{A} (\bm{x}, t) \cdot d \bm{x} \ = \ 
 \int \bm{B} (\bm{x}, t) \cdot d \bm{S} \ = \ e \,\,\Phi (t).
\end{equation}
However, their analysis
is still lacking a convincing explanation about where and how the cancellation
argument by Singleton et al. is not justified.
To our belief, more orthodox way to answer the above question is to go
back to the basic equation of quantum mechanics, i.e. the Schr\"{o}dinger
equation of a charged particle under the presence of time-dependent
magnetic flux of the solenoid as faithful as possible to the basic principle
of quantum mechanics. (We emphasize that a wave function in quantum mechanics
is a complex number, which necessarily contains the information of phase,
which is essential in the physics of the AB-phase.)
This analysis will be performed in the next section.

\section{\label{sec3}Quantum mechanics of the time-dependent AB-phase shift}

The basic lagrangian describing the motion of the electron under the influence of the 
external electromagnetic potential is given by
\begin{eqnarray}
 L \ = \ \frac{1}{2} \,m \,\dot{\bm{x}}^2 \ + \ e \,
 \left[ \,A^0 (\bm{x}, t) - \dot{\bm{x}} \cdot \bm{A} (\bm{x}, t) \right] ,
\end{eqnarray}
where $A^\mu (\bm{x}, t) = ( A^0 (\bm{x}, t), \bm{A} (\bm{x}, t))$ represents the
4-vector potential generated by externally given charge and current densities.
The equation  of motion for the electron derived from the above Lagrangian
is given by
\begin{eqnarray}
 m \,\ddot{\bm{x}} \ = \ - \,e \,\left[ \bm{E} \ + \ 
 \dot{\bm{x}} \times \bm{B} \right] , \label{Eq:equation_of_motion}
\end{eqnarray}
with
\begin{eqnarray}
 \bm{E} \ = \ - \,\nabla A^0 \ - \ 
 \frac{\partial}{\partial t} \,\bm{A}, \ \ \ \ 
 \bm{B} \ = \ \nabla \times \bm{A} . \label{Eq:Electric_magnetic_field}
\end{eqnarray}
The corresponding Schr\"{o}dinger equation for the electron is given by
\begin{eqnarray}
 i \,\frac{\partial}{\partial t} \,\psi = 
 \left[\, - \,\frac{1}{2 \,m} 
 \left( \nabla + i \,e \,\bm{A} (\bm{x}, t) \right)^2
 - e \,A^0 (\bm{x}, t) \right] \psi . \hspace{4mm}
\end{eqnarray}
If we carry out the gauge transformation
\begin{eqnarray}
 A_\mu (x) &\rightarrow& A^{\prime}_\mu (x) \ = \ A_\mu (x) \ - \  
 \partial_\mu \chi (x) , \\
 \psi (x) &\rightarrow& \psi^\prime (x) \ = \ 
 e^{\,- \,i \,e \chi (x)} \,\psi (x) ,
\end{eqnarray} 
the new electron wave function $\psi^\prime (x)$ is shown to obey the following 
Schr\"{o}dinger equation
\begin{eqnarray}
 i \,\frac{\partial}{\partial t} \,\psi^\prime = 
 \left[\, - \,\frac{1}{2 \,m} 
 \left( \nabla + i \,e \,\bm{A}^\prime (\bm{x}, t) \right)^2
 - e \,A^{\prime 0} (\bm{x}, t) \,\right] \,\psi^\prime. \hspace{6mm}
\end{eqnarray}
Interestingly, we can formally eliminate the vector potential by making use
of a gauge transformation.
In fact, let us define the new electron wave function $\tilde{\psi} (\bm{x}, t)$ by
\begin{equation}
 \psi (\bm{x}, t) \ \equiv \ e^{\,i \,e \,\Lambda (x)} \,
 \tilde{\psi} (\bm{x}, t) , 
\end{equation} 
with 
\begin{equation} 
 \Lambda (x) \ = \ - \,\int^{x^\mu}_{x_0^\mu} \,
 A_\mu (x^\prime) \,d x^{\prime \mu} , \label{Eq:gauge_function_Lambda}
\end{equation}
where, $x^\mu_0$ is supposed to represent some fixed reference point in the
4-dimensional Minkowski space.
Here we must keep in mind the fact that the line-integral for defining the 
function $\Lambda (x)$ generally depends on the choice of the space-time 
path $C$ connecting $x^\mu_0$ and $x^\mu$.
Here, by using 
\begin{eqnarray}
 \nabla \,\Lambda (x) &=& \ \ \nabla \int^{\bm{x}}_{\bm{x}_0} \,
 \bm{A} (\bm{x}^\prime, t) \cdot d \bm{x}^\prime 
  \ = \ \bm{A} (\bm{x}, t) , \\
 \frac{\partial}{\partial t} \,\Lambda (x)
 &=& - \,\frac{\partial}{\partial t}
 \,\int^t_{t_0} \,A^0 (\bm{x}, t^\prime) \,d t^\prime
 \ = \ - \,A^0 (\bm{x}, t) , \hspace{5mm}
\end{eqnarray}
it can be shown that  $\tilde{\psi} (\bm{x}, t)$  satisfies the following free 
Schr\"{o}dinger eq. 
\begin{eqnarray}
 i \,\,\frac{\partial}{\partial t} \,\tilde{\psi} (\bm{x}, t) \ = \ 
 - \,\frac{1}{2 \,m} \,\nabla^2 \,\tilde{\psi} (\bm{x}, t) .
\end{eqnarray}
A critically important point is that, in an appropriately  
chosen {\it simply-connected} space-time region corresponding to our problem, 
the gauge function  $\Lambda (x)$ can be defined 
as a unique function of space-time coordinate  $x^\mu$, that is,
independently on the space-time path $C$ connecting
$x^\mu_0$ and $x^\mu$. 
(A simply-connected region here means the space-time region, which
never contains the solenoid in it.
The above-mentioned property is vitally important for that 
the whole argument in the present paper is meaningful. 
Since its proof needs very careful consideration, we discuss it in Appendix A.)

\vspace{3mm}
To sum up, for a simply connected space-time region which we can
choose in our problem, the solution of the 
Schr\"{o}dinger equation
 \begin{eqnarray}
 i \,\frac{\partial}{\partial t} \,\psi = 
 \left[ - \,\frac{1}{2 \,m} 
 \left( \nabla + i \,e \,\bm{A} (\bm{x}, t) \right)^2
 - e \,A^0 (\bm{x}, t) \right]  \psi. \hspace{5mm}
\end{eqnarray} 
can be expressed as
\begin{eqnarray}
 \psi (\bm{x}, t) \ = \ e^{\,i \,e \,\Lambda (x)} \,
 \tilde{\psi} (\bm{x}, t) ,
\end{eqnarray}
with
\begin{eqnarray} 
 \Lambda (x) \ = \ - \,\int^{x^\mu}_{x_0^\mu} \,
 A_\mu (x^\prime) \,d x^{\prime \mu}
\end{eqnarray}
where $\tilde{\psi} (\bm{x}, t)$ is a solution of {\it free} Schr\"{o}dinger equation.

\vspace{4mm}
\begin{figure}[ht]
\begin{center}
\includegraphics[width=7.0cm]{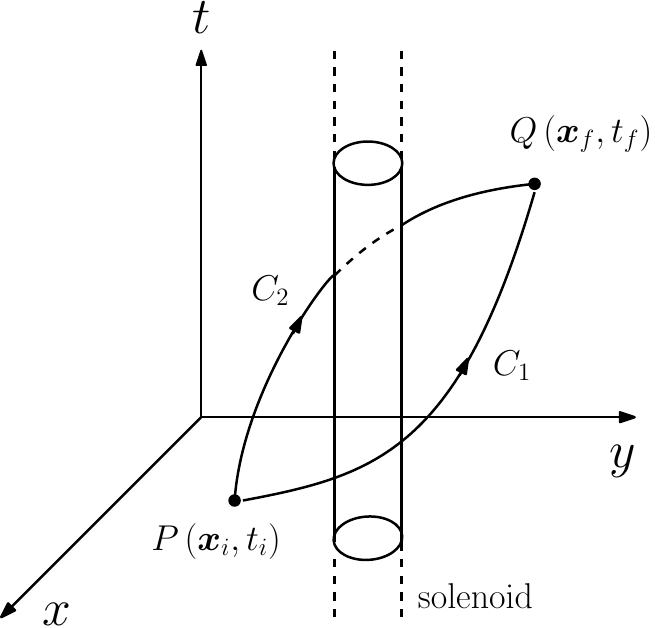}
\caption{Two paths $C_1$ and $C_2$ in the Minkowski space, which start  
from the space-time point $P (\bm{x}_i, t_i)$ and ends at $Q (\bm{x}_f, t_f)$.}
\label{Fig:time-dep.B_space-time}
\end{center}
\end{figure}

This implies that the phase change of the electron wave function 
along the path  $C_1$  (from space-time point $P$ to $Q$) illustrated
in Fig.\ref{Fig:time-dep.B_space-time} is given by
\begin{eqnarray}
 \Delta \phi_{AB} (C_1) \ = \ - \,
 \int_{C_1} \,A_\mu (x^\prime) \,d x^{\prime \mu} .
\end{eqnarray}
Similarly, the phase change along the path $C_2$ is given by
\begin{eqnarray}
 \Delta \phi_{AB} (C_2) \ = \ - \,
 \int_{C_2} \,A_\mu (x^\prime) \,d x^{\prime \mu} .
\end{eqnarray}
In view of the ambiguity of the vector potential, one may consider the freedom
of further gauge transformation
\begin{eqnarray}
 A_\mu (x) \ \rightarrow \ A_\mu^\prime (x) \ = \ 
 A_\mu (x) \ - \ \partial_\mu \chi (x)
\end{eqnarray}
with  $\chi (x)$ being an {\it arbitrary scalar function}.
Under the gauge transformation,   $\Delta \phi_{AB} (C_1)$ and $\Delta \phi_{AB} (C_2)$
change as
\begin{eqnarray}
 \Delta \phi_{AB} (C_1) &\rightarrow& - \,e \,
 \left\{\,\int_{C_1} \!\!A_\mu (x^\prime) \,d x^{\prime \mu}
 \ - \ \int_{C_1} \! \partial_\mu \chi (x^\prime) \,
 d x^{\prime \mu} \,\right\} \nonumber \\
 &=& - \,e \int_{C_1}\!\! A_\mu (x^\prime) \,d x^{\prime \mu}
 \ + \ e \,\left[\, \chi (Q) - \chi (P) \,\right] , \hspace{6mm}
\end{eqnarray}
and
\begin{eqnarray}
 \Delta \phi_{AB} (C_2) &\rightarrow&
 - \,e \int_{C_2} \!\!A_\mu (x^\prime) \,d x^{\prime \mu}
 \ + \ e \left[\, \chi (Q) - \chi (P) \,\right] \!. \hspace{5mm}
\end{eqnarray}
Observable AB-phase shift is given by the phase difference between the two paths,
\begin{eqnarray}
 \phi_{AB} \ = \ \Delta \phi_{AB} (C_1) \ - \ 
 \Delta \phi_{AB} (C_2)
\end{eqnarray}
As long as the gauge function $\chi (x)$ is {\it single-valued} (not multi-valued), the
gauge freedom parts just cancel out, thereby being led to
\begin{eqnarray}
 \phi_{AB} &=& - \,e \,\int_{C_1} \,A_\mu (x^\prime) \,d x^{\prime \mu}
 \ + \ e \,\int_{C_2} \,A_\mu (x^\prime) \,d x^{\prime \mu} 
 \ = \ - \,e \,\oint_{C_1 - C_2} \,A_\mu (x^\prime) \,d x^{\prime \mu} .
\end{eqnarray}
This precisely reproduces the 1st form of the covariant expression given 
by Eq.(\ref{Eq:1st_form}).

\vspace{3mm}
At this point, we recall that equation of motion for the charged particle is 
given by Eq.(\ref{Eq:equation_of_motion}).
Note that the electromagnetic fields appearing in this equation are generated 
according to the Maxwell equation
\begin{eqnarray}
 \partial_\mu F^{\mu \nu} \ = \ j^\nu ,
\end{eqnarray}
where $j^\nu (x)$ is the external source current provided by the solenoid.
In terms of the vector potential, the above Maxwell equation can be expressed as
\begin{eqnarray}
 \square \,A^\nu \ - \ \partial^\nu \,( \partial_\mu A^\mu )
 \ = \ j^\nu .
\end{eqnarray}
As is widely known, the vector potential has gauge ambiguity.
In our time-dependent problem, it is most convenient to start with 
the choice of the {\it Lorenz gauge} : 
\begin{eqnarray}
 \partial_\mu A^\mu \ = \ 0 .
\end{eqnarray}
In the Lorenz gauge, the 4-vector potential satisfies a very  simple equation
\begin{eqnarray}
 \square \,A^\nu \ = \ j^\nu .
\end{eqnarray}
Its solution is well-known and given in the form
\begin{eqnarray}
 A^\nu (\bm{x}, t) \ = \ \frac{1}{4 \,\pi} \,\int \,
 \frac{j^\nu (\bm{x}^\prime, t_R)}
 {\vert \bm{x} - \bm{x}^\prime \vert} \,d^3 x^\prime
\end{eqnarray}
where  $t_R \equiv t - \vert \bm{x} - \bm{x}^\prime \vert / c$
is the so-called {\it retardation time}. 
(Although we use the natural unit, we retain
the speed of light $c$ only in this expression in order to emphasize the
retardation time is controlled by the light speed $c$.)
To be more precise, even after Lorenz gauge fixing, there is a freedom of residual 
gauge transformation represented as
\begin{eqnarray}
 A^\nu (\bm{x}, t) \ = \ \frac{1}{4 \,\pi} \,\int \,
 \frac{j^\nu (\bm{x}^\prime, t_R)}
 {\vert \bm{x} - \bm{x}^\prime \vert} \,d^3 x^\prime \label{Eq:general_form_A}
 \ + \ \partial^\nu \chi . \hspace{5mm}
\end{eqnarray}
We say it the {\it residual gauge transformation}, because the Lorenz gauge condition 
$\partial_\mu A^\mu = 0$ is retained as long as the gauge function  $\chi$ is
chosen to satisfy the condition : 
\begin{eqnarray}
 \square \,\chi = 0 .
\end{eqnarray}
Note however that, since the external gauge potential  $A^\nu$ can be treated as 
classical field in the present problem, we do not need to confine to such a 
residual gauge transformation in the Lorenz gauge.
By allowing {\it  any choice} of  $\chi$, the vector potential in arbitrary gauge
can be expressed in the form given by Eq.(\ref{Eq:general_form_A}). 
In fact, in classical gauge theory,
the explicit form of gauge transformation from Lorenz to Coulomb and other gauges
are known \cite{Jackson2002}. 
(One might suspect that the situation might change drastically, if one allows multi-valued 
gauge transformations \cite{BL1978}. 
In fact, in the case of the standard time-independent
AB-effect, there were controversial dispute  about the existence of the AB-effect itself.
At the end, however, it turned out that the physics does not change at all even by 
multi-valued gauge transformations \cite{Waka2024,Kreizschmar1965, BB1983, BMM2016}. 
Here, we take it for granted that it would be also true for the time-dependent AB-effect, 
thereby confining to regular or single-valued gauge transformations.) 

%

\vspace{3mm}
In any case, from the argument above, we conclude that, in any single-valued 
gauge, the AB-phase shift for a closed space-time path can be expressed as
\begin{eqnarray}
 \phi_{AB} \,= \, 
 - \,e \,\oint_{C_1 - C_2} \,A_\mu (x^\prime) \,d x^{\prime \mu}
\end{eqnarray}
where
\begin{eqnarray}
 A^\nu (\bm{x}, t) \ = \ \frac{1}{4 \,\pi} \,\int \,
 \frac{j^\nu (\bm{x}^\prime, t_R)}
 {\vert \bm{x} - \bm{x}^\prime \vert} \,d^3 x^\prime
 \ + \ \partial^\nu \chi , \hspace{5mm}
\end{eqnarray}
with  $t_R \equiv t - \vert \bm{x} - \bm{x}^\prime \vert / c$. 
If the motion of charged particle is non-relativistic, we can neglect the difference between
$t_R$ and $t$, so that we get
\begin{eqnarray*}
 A^\nu (\bm{x}, t) \ \simeq \ \frac{1}{4 \,\pi} \,\int \,
 \frac{j^\nu (\bm{x}^\prime, t)}
 {\vert \bm{x} - \bm{x}^\prime \vert} \,d^3 x^\prime
 \ + \ \partial^\nu \chi .
\end{eqnarray*}
Or, in the 3-vector notation with  $j^\nu (\bm{x}, t) = ( \rho (\bm{x}, t), \bm{j} (\bm{x}, t))$,
we have 
\begin{eqnarray}
 A^0 (\bm{x}, t) &\simeq& \frac{1}{4 \,\pi} \, \int \,
 \frac{\rho (\bm{x}^\prime, t)}{\vert \bm{x} - \bm{x}^\prime \vert}
 \,d^3 x^\prime  
 \ + \ \frac{\partial}{\partial t} \,\chi (\bm{x}, t) , \hspace{5mm} \\
 \bm{A}(\bm{x}, t) &\simeq& \frac{1}{4 \,\pi} \, \int \,
 \frac{\bm{j} (\bm{x}^\prime, t)}{\vert \bm{x} - \bm{x}^\prime \vert}
 \,d^3 x^\prime  
 \ - \ \nabla \chi (\bm{x}, t) .
\end{eqnarray}
%

Now we consider our concrete setting of infinitely-long solenoid with radius $R$.
The charge and current densities of such a solenoid are given 
by \cite{WKZZ2018,Waka2024}
\begin{eqnarray}
 \rho (\bm{x}, t) &=& 0 ,\\
 \bm{j} (\bm{x}, t) &=& \Phi (t) \,
 \delta (\rho - R) \,
 \bm{e}_\phi \ \ \ \ \ (\rho = \sqrt{x^2 + y^2}) . \hspace{5mm}
\end{eqnarray}
This gives
\begin{eqnarray}
 A^0 (\bm{x}, t) \ = \ 0, \ \ \ 
 \bm{A}(\bm{x}, t) \ = \ \bm{A}^{(S)} (\bm{x}, t) \hspace{6mm}
\end{eqnarray}
with
\begin{eqnarray}
 \bm{A}^{(S)} (\bm{x}, t) \ = \ 
 \left\{ \begin{array}{cc}
 \frac{\rho \,\Phi (t)}{2 \,\pi \,R^2} \,
 \bm{e}_\phi \ & \ \ \ (\rho < R)  , \\
 \frac{\Phi (t)}{2 \,\pi \rho} \,
 \bm{e}_\phi \ & \ \ \ (\rho \geq R) . \\
 \end{array} \right. \hspace{5mm}
\end{eqnarray}
As we have already discussed, as long as we confine to regular or 
single-valued gauge transformations, the AB-phase shift for a closed path
is not affected by the arbitrariness of the gauge function $\chi$.  
The prediction for the AB-phase shift is therefore given by
\begin{eqnarray}
 \phi_{AB} \ = \ - \,e \,\left[ \,
 \oint_{C_1 - C_2} \!\! A^0 (\bm{x}, t) \,d t -  
 \oint_{C_1 - C_2} \!\! \bm{A} (\bm{x}, t) \cdot d \bm{x} \,
 \right] 
 \ = \  e \,\oint_{C_1 - C_2} \!\!\bm{A} (\bm{x}, t) \cdot d \bm{x}. 
 \label{Eq:Stokes3_1}
\end{eqnarray}
It seems that everything is going right so far.
From this last expression, several authors proceed as follows.
(See the paper \cite{JZWLD2017} by Jing et al., for example.)
Namely, they naively assume that the Stokes theorem of the following
3-dimensional form holds :
\begin{eqnarray}
 \oint_{C_1 - C_2} \bm{A} (\bm{x}, t) \cdot d \bm{x} &=&
 \int \,(\nabla \times \bm{A} (\bm{x}, t)) \cdot d \bm{S} 
 \ = \ 
 \int \bm{B} (\bm{x}, t) \cdot d \bm{S} . \label{Eq:Stokes3_2}
\end{eqnarray}
If it were correct, it would give the following prediction for the 
time-dependent AB-phase shift,
\begin{equation}
 \phi_{AB} \ = \ e \,\int \bm{B} (\bm{x}, t) \cdot d \bm{S} \ = \ e \,\, \Phi (t) \ \ ?
 \label{Eq:Stokes3_3}
\end{equation}
Somewhat unexpectedly, we realize that the above manipulation is {\it not} actually 
justified if the vector potential $\bm{A}$ is time-dependent. 
The reason is because, on the space-time curves $C_1$ and $C_2$, 
spatial coordinate variable $\bm{x}$ and the time variable $t$ cannot be 
treated as independent integration variables.
To understand this subtlety, we find it mandatory to reanalyze 
the 4-dimensional Stokes theorem by paying the greatest care. Besides,
an important byproduct of this study is more transparent understanding
about the validity or invalidity of the cancellation argument by Singleton et al.
between the time-dependent magnetic field and the induced electric field.

\section{\label{sec4}On the treatment based on the 4-dimensional Stokes theorem}

The 4-dimensional Stokes theorem was already written down 
in (\ref{Eq:4-dim_Stokes_theorem}). However, to understand more precise meaning 
of this important theorem, it is better to express it in more accurate form as follows :  
\begin{equation}
 \oint_{C_1 - C_2} A_\mu (x) \,d x^\mu \ = \ \frac{1}{2} \,
 \iint_{S \, (C_1 - C_2)} F_{\mu \nu} (x) \, d x^\mu \wedge d x^\nu  . 
 \label{Eq:4-dim_ST}
\end{equation}
Here, the l.h.s of the above equation represents the line-integral of the 4-vector
potential $A_\mu (x)$ along the closed space-time path $C_1 - C_2$ as illustrated
in Fig. \ref{Fig:time-dep.B_space-time}. On the other hand, the r.h.s. represents
the integral of the field-strength tensor $F_{\mu \nu} (x)$ over the
2-dimensional surface $S$, the boundary of which is specified by the 
closed path $C_1 - C_2$ in the Minkowski space.  

\vspace{2mm}
To correctly understand the meaning of the equality (\ref{Eq:4-dim_ST}),
we consider our concrete situation in which the infinitely-long solenoid 
with the radius $R$ directed to 
the $z$-axis is placed at the origin of $xy$-plane, and the space-time path $C_1$
and $C_2$ connecting the two space-time points $(\bm{x}_i, t_i)$ and $(\bm{x}_f, t_f)$
are parametrized as follows. Namely, suppose that the path $C_1$ is given by
a circular arc with radius $\rho_0 \,(> R)$ around the coordinate origin, which is
taken to be the center of the solenoid. 
Then, the space-time points on the path $C_1 (t_i \rightarrow t_f)$ can be  
parametrized as
\begin{equation}
  (t (\phi), x (\phi), y (\phi), z (\phi)) =  
  (f (\phi), \rho_0 \cos \phi, \,\rho_0 \sin \phi, \,0) , \label{Eq:C_1} \hspace{5mm}
\end{equation}
where $\phi_i \leq \phi \leq \phi_f$ with $\phi_i \equiv \phi (t_i) $ and
$\phi_f \equiv \phi (t_f)$. Here, it is assumed that $\rho_0$ is larger than the radius $R$
of the solenoid. Similarly, the space-time points on the path 
$- \,C_2 (t_f \rightarrow t_i)$ are parametrized as
\begin{equation}
 (t (\phi), x (\phi), y (\phi), z (\phi)) = 
 (g (\phi), \,\rho_0 \cos \phi, \,\rho_0 \sin \phi, \,0),  \label{Eq:-C_2}
\end{equation}
where $\phi_f  \leq \phi \leq \phi_i + 2 \,\pi$ with $\phi_f \equiv \phi (t_f) $ and
$\phi_i \equiv \phi (t_i)$.
In the above parametrizations, the functions
$f (\phi)$ and $g(\phi)$ are supposed to satisfy the conditions
\begin{equation}
 f (\phi_i) \ = \ t_i, \ \ f(\phi_f) \ = \ t_f,
\end{equation}
and
\begin{equation}  
 g(\phi_f) \ = \ t_f, \ \ g(\phi_i + 2 \,\pi) \ = \ t_i. 
\end{equation}
We also need to parametrize the space-time points on the 2-dimensional
surface $S$ surrounded by $C_1 - C_2$. We choose to take as
\begin{equation}
 t  (\rho, \phi) \ = \ \left\{ \begin{array}{cl}
 F (\rho) \,f (\phi) \ \  &  \ \ (\phi_i \leq \phi \leq \phi_f), \\
 G (\rho) \ g (\phi) \ \  &  \ \ (\phi_f  \leq \phi \leq \phi_i + 2 \,\pi), \\
 \end{array} \right.  
\end{equation}
and
\begin{equation}
 x (\rho, \phi) = \rho \cos \phi, \ \ 
 y (\rho, \phi) =  \rho \sin \phi, \  \ 
 z (\rho, \phi) = 0 , 
\end{equation}
with $\phi_i  \leq \phi \leq \phi_i + 2 \,\pi$.
Here, $F (\rho)$ and $G (\rho)$ with $0 \leq \rho \leq \rho_0$ are any functions
for parametrizing  points on the 2-dimensional surface surrounded by the 
boundary $C_1 - C_2$, where $C_1$ and $- \,C_2$ are respectively specified by 
Eq.(\ref{Eq:C_1}) and Eq.(\ref{Eq:-C_2}). Here, the functions 
$F (\rho)$ and $ G (\phi)$ are supposed to satisfy the following conditions
\begin{equation}
 F (\rho_0) = 1, \ \ \ G (\rho_0) = 1.
\end{equation}  
This is because the space-time points 
$(t (\rho_0, \phi), x (\rho_0, \phi), y (\rho_0, \phi), z (\rho_0, \phi))$ must lie 
on the curves $C_1$ and $- \,C_2$. 

\vspace{2mm}
To evaluate the surface-integral of the field-strength tensor appearing in the
4-dimensional Stokes theorem, we need to specify the integration measure
over the surface $S$.
The differentials of the space-time coordinates on $S$ are given by
\begin{equation}
 d t (\rho, \phi) = \left \{ \begin{array}{ll}
 F^\prime (\rho) \,f (\phi) \,d \rho + F (\rho) \,f^\prime (\phi) \,d \phi \ \ & \ \  
  (\phi_i \leq \phi \leq \phi_f) , \\
 G^\prime (\rho) \,g (\phi) \,d \rho + G (\rho) \,g^\prime (\phi) \,d \phi \ \ & \ \  
 (\phi_f  \leq \phi \leq \phi_i + 2 \,\pi) .\\
 \end{array} \right.
\end{equation}
and by
\begin{eqnarray}
 d x (\rho, \phi) &=& \cos \phi \,d \rho \ - \ \rho \,\sin \phi \,d \phi \nonumber \\
 d y (\rho, \phi) &=& \sin \phi \,d \rho \ + \ \rho \,\cos \phi \,d \phi \hspace{8mm} \\
 d z (\rho, \phi) &= & 0. \nonumber
\end{eqnarray}
The integration measure for the surface integral is therefore given as follows.
First, for $\phi_i \leq \phi \leq \phi_f$, we get
\begin{eqnarray}
 d t (\rho, \phi) \wedge d x (\rho, \phi) 
 &=& (\,F^\prime (\rho) f (\phi) \,d \rho + F (\rho) f^\prime (\phi) \,d \phi \,)
 \wedge (\cos \phi \,d \rho - \rho \sin \phi \,d \phi ) \hspace{8mm} 
 \nonumber \\
 &=& \left\{ \,- \,\rho \,F^\prime (\rho) f (\phi) \sin \phi -  
 F (\rho) f^\prime (\phi) \cos \phi \,\right\} d \rho \wedge d \phi,  \\
 d t (\rho, \phi) \wedge d y (\rho, \phi) 
 &=&  (\,F^\prime (\rho) f (\phi) d \rho + F (\rho) f^\prime (\phi) \,d \phi )
 \wedge (\sin \phi \,d \rho + \rho \cos \phi \,d \phi ) \nonumber \\
 &=& \left\{\, + \,\rho \,F^\prime (\rho) \,f (\phi) \,\cos \phi - 
 F (\rho) \,f^\prime (\phi) \,\sin \phi \,\right\} \,d \rho \wedge d \phi, \\
 d t (\rho, \phi) \wedge d z (\rho, \phi)  &=& 0.
\end{eqnarray}
Similarly, for $\phi_f  \leq \phi \leq \phi_i + 2 \,\pi$, we obtain
\begin{eqnarray}
 d t (\rho, \phi) \wedge d x (\rho, \phi) 
 &=&  \left\{\, - \,\rho \,G^\prime (\rho) g (\phi) \sin \phi  -  
 G (\rho) g^\prime (\phi) \cos \phi \,\right\} d \rho \wedge d \phi, \\
 d t (\rho, \phi) \wedge d y (\rho, \phi) 
 &=& \left\{\, + \,\rho \,G^\prime (\rho) g (\phi) \,\cos \phi - 
 G (\rho) g^\prime (\phi) \,\sin \phi \,\right\} d \rho \wedge d \phi, \hspace{8mm} \\
 d t (\rho, \phi) \wedge d z (\rho, \phi)  &=&  0.  
\end{eqnarray}
Based on these preparations, we are now ready to evaluate the contribution of the magnetic
field and the induced electric field to the surface integral appearing in the 4-dimensional
Stokes theorem. 

\subsection{Electric field contribution}

Let us begin with the contribution of the electric field.
We first recall that the induced electric field in our setting is given as
\begin{equation}
 \bm{E} \ = \ - \,\frac{\partial \bm{A}}{\partial t} \ = \ \left\{ \begin{array}{cc}
 - \,\frac{1}{2} \,\dot{B} (t) \,\rho \,\,\,\bm{e}_\phi \ \ & \ \ (\rho < R)  \\
 - \,\frac{R^2}{2} \,\dot{B} (t) \,\frac{1}{\rho} \,\,\bm{e}_\phi \ \ & \ \ (\rho \geq R) \\
 \end{array} \right. ,
\end{equation}
with $\dot{B} (t) \equiv \frac{d B (t)}{d t}$. The corresponding field strength
tensors are given by
\begin{equation}
 F_{0x} = \frac{1}{2} \,\dot {B} (t) \,\rho \sin \phi, \  \ 
 F_{0y} = - \,\frac{1}{2} \,\dot{B} (t) \,\rho \cos \phi, \ \    
 F_{0z} = 0, 
\end{equation}
for $\rho < R$, and 
\begin{equation}
 F_{0x} = \frac{R^2}{2} \,\dot {B} (t) \,\frac{1}{\rho} \sin \phi, \ \  
 F_{0y} = \- \,\frac{R^2}{2} \,\dot{B} (t) \,\frac{1}{\rho} \cos \phi, \ \  
 F_{0z} = 0,
\end{equation}
for $\rho > R$.
Our next task is to evaluate the contribution of the induced electric field to the
quantity
\begin{equation}
 \frac{1}{2} \,F_{\mu \nu} \,d x^\mu \wedge d x^\nu.
\end{equation}
It is carried out by dividing the 2-dimensional surface $S$ into the following 
4 pieces,
\begin{eqnarray*}
 &\,& (\rm{i}) \ \ \ \phi_i \leq \phi \leq \phi_f, \ \ \ 0 \leq \rho < R, \\
 &\,& (\rm{ii}) \ \ \phi_i \leq \phi \leq \phi_f,\ \ R \leq \rho \leq \rho_0, \\
 &\,& (\rm{iii}) \ \ \phi_f \leq \phi \leq \phi_i + 2 \,\pi, \ \ 0 \leq \rho < R, \\ 
 &\,& (\rm{iv}) \ \ \phi_f \leq \phi \leq \phi_i + 2 \,\pi, \ \ R \leq \rho < \rho_0.
 \hspace{16mm}
\end{eqnarray*}

\vspace{2mm}
\noindent
(i) contribution from the region $\phi_i \leq \phi \leq \phi_f, \  0 \leq \rho < R$

\vspace{2mm}
In this region, the contribution of the electric field to $\frac{1}{2} \,F_{\mu \nu} \,
d x^\mu \wedge d x^\nu$ is given by
\begin{eqnarray}
 &\,& \left. \frac{1}{2} \,F_{\mu \nu} \,d x^\mu \wedge d x^\nu \,\right|_{electric}
 \ = \ F_{0x} \,d t \wedge d x  +  F_{0y} \,d t \wedge d y  +  
 F_{0z} \,d t \wedge d z \nonumber \\
 &\,& \ = \ \frac{1}{2} \,\dot{B} (t) \,\rho \,\sin \phi 
 \times \left\{\, - \,\rho \,F^\prime (\rho) \,f (\phi) \sin \phi \ - 
 \ F (\rho) \,f^\prime (\phi) \cos \phi \,\right\} \,d \rho \wedge d \phi \nonumber \\
 &\,& \ \ - \ \frac{1}{2} \,\dot{B} (t) \,\rho \cos \phi 
 \times \left\{\, + \,\rho \,F^\prime (\rho) \,f (\phi) \cos \phi \ - \ 
 F (\rho) \,f^\prime (\phi) \sin \phi \,\right\} \, d \rho \wedge d \phi 
 \hspace{8mm} \nonumber \\
 &\,& \ = \ \frac{1}{2} \,\dot{B} (t) \,\rho^2 \,F^\prime (\rho) \,f (\phi) \,d \rho \wedge d \phi. 
\end{eqnarray}
For convenience, we introduce an auxiliary function $\tilde{\cal B} (\rho, \phi)$ defined by
\begin{equation}
 \tilde{\cal B} (\rho, \phi) \ \equiv \ B (F (\rho) \,f (\phi)). \label{Btilde}
\end{equation}
Note that it satisfies the following equation
\begin{equation}
 \left. \frac{\partial \tilde{\cal B} (\rho, \phi)}{\partial \rho} \ = \ 
 F^\prime (\rho) \,f (\phi) \,\dot{B} (t) \,\right\vert_{t = F (\rho) \,f (\phi)} .
\end{equation}
Using this relation, we find that the electric field contribution from the region (i)
can be expressed as
\begin{equation}
 \left. \frac{1}{2} \,F_{\mu \nu} \,d x^\mu \wedge d x^\nu \,\right\vert_{electric}
 = - \,\frac{1}{2} \,\,\rho^2 \,\,\frac{\partial \tilde{\cal B} (\rho, \phi)}{\partial \rho} \,
 d \rho \wedge d \phi .
\end{equation}
Integrating it over $\rho$ from $0$ to $R$, we get
\begin{eqnarray}
 &\,& - \,\int_0^R \,d \rho \,\,\frac{\rho^2}{2} \,
 \frac{\partial \tilde{\cal B} (\rho, \phi)}{\partial \rho}
 \ = \ - \,\frac{R^2}{2} \,\tilde{\cal B} (R, \phi) \ + \ 
 \int_0^R \,d \rho \,\rho \,\tilde{\cal B} (\rho, \phi). \hspace{8mm}
\end{eqnarray}
In this way, we eventually find that the electric field contribution to the surface
integral from the region (i)  is given by
\begin{eqnarray}
 \left. \iint_{region \ \rm{(i)}} \,\frac{1}{2} \,F_{\mu \nu} \,d x^\mu \wedge d x^\nu
 \,\right\vert_{electric}
 \ = \ \int_{\phi_i}^{\phi_f} d \phi \,\left[
 - \,\frac{R^2}{2} \,\tilde{\cal B} (R, \phi) + \int_0^R \,d \rho \,\rho \,
 \tilde{\cal B} (\rho, \phi) \right] . 
\end{eqnarray}

\vspace{2mm}
\noindent
(ii) contribution from the region $\phi_i \leq \phi \leq \phi_f, R \leq \rho \leq \rho_0$

\vspace{2mm}
The contribution of the electric field to $\frac{1}{2} \,F_{\mu \nu} \,
d x^\mu \wedge d x^\nu$ from this region is given by
\begin{eqnarray}
 &\,& \left. \frac{1}{2} \,F_{\mu \nu} \,d x^\mu \wedge d x^\nu \,\right|_{electric}
 \nonumber \\
 &\,& = F_{0x} \,d t \wedge d x  +  F_{0y} \,d t \wedge d y  +  
 F_{0z} \,d t \wedge d z \nonumber \\
 &\,& = \frac{R^2}{2} \,\dot{B} (t) \,\frac{1}{\rho} \,\sin \phi 
 \times \left\{\, - \,\rho \,F^\prime (\rho) \,f (\phi) \sin \phi \ - 
 \ F (\rho) \,f^\prime (\phi) \cos \phi \,\right\} \,d \rho \wedge d \phi 
 \hspace{12mm} \nonumber \\
 &\,& \ - \frac{R^2}{2} \,\dot{B} (t) \,\frac{1}{\rho} \,\cos \phi
 \times \left\{\, + \,\rho \,F^\prime (\rho) \,f (\phi) \cos \phi \ - \ 
 F (\rho) \,f^\prime (\phi) \sin \phi \,\right\} \, d \rho \wedge d \phi 
 \hspace{12mm} \nonumber \\
 &\,&  = - \,\frac{R^2}{2} \,\dot{B} (t) \,F^\prime (\rho) \,f (\phi) \,d \rho \wedge d \phi
 \ = \ - \,\frac{R^2}{2} \,\frac{\partial \tilde{\cal B} (\rho, \phi)}{\partial \rho} \,
 d \rho \wedge d \phi .
\end{eqnarray}
Integrating over $\rho$ from $R$ to $\rho_0$, this gives
\begin{eqnarray}
 - \,\int_{R}^{\rho_0} \,d \rho \,\,\frac{R^2}{2} \,
 \frac{\partial \tilde{\cal B} (\rho, \phi)}{\partial \rho}
 \ = \  
 - \,\frac{R^2}{2} \,\tilde{\cal B} (\rho_0, \phi) \ + \ 
 \frac{R^2}{2} \,\tilde{\cal B} (R, \phi) .  \hspace{8mm}
\end{eqnarray}
Thus, the electric field contribution from the region (ii) is given by
\begin{eqnarray}
 \left. \iint_{region \ \rm{(ii)}} \,\frac{1}{2} \,F_{\mu \nu} \,d x^\mu \wedge d x^\nu
 \,\right\vert_{electric} 
 \ = \ \int_{\phi_i}^{\phi_f} \,d \phi \,\left[\,
  - \,\frac{R^2}{2} \,\tilde{\cal B} (\rho_0, \phi)  +  
 \frac{R^2}{2} \,\tilde{\cal B} (R, \phi) \,\right] .  \hspace{5mm}
\end{eqnarray}

\vspace{2mm}
\noindent
(iii) contribution from the region $\phi_f \leq \phi \leq \phi_i + 2 \,\pi, \ 0 \leq \rho < R$

\vspace{2mm}
Similarly, the contribution of the electric field to 
$\frac{1}{2} \,F_{\mu \nu} \, d x^\mu \wedge d x^\nu$ from the region (iii) is given by
\begin{eqnarray}
 &\,& \left. \frac{1}{2} \,F_{\mu \nu} \,d x^\mu \wedge d x^\nu \,\right|_{electric}
 \nonumber \\
 &\,& = F_{0x} \,d t \wedge d x \ + \ F_{0y} \,d t \wedge d y \ + \ 
 F_{0z} \,d t \wedge d z \nonumber \\
 &\,& = \frac{1}{2} \,\dot{B} (t) \,\rho \,\sin \phi
 \times \left\{\, - \,\rho \,G^\prime (\rho) \,g (\phi) \sin \phi \ - 
 \ G (\rho) \,g^\prime (\phi) \cos \phi \,\right\} \,d \rho \wedge d \phi \nonumber \\
 &\,& \ - \frac{1}{2} \,\dot{B} (t) \,\rho \,\cos \phi
 \times \left\{\, + \,\rho \,G^\prime (\rho) \,g (\phi) \cos \phi \ - \ 
 G (\rho) \,g^\prime (\phi) \sin \phi \,\right\} \, d \rho \wedge d \phi 
 \hspace{12mm} \nonumber \\
 &\,& = - \,\frac{1}{2} \,\dot{B} (t) \,\rho^2 \,G^\prime (\rho) \,g (\phi) \,d \rho \wedge d \phi
 \ = \ - \,\frac{\rho^2}{2} \,\frac{\partial \hat{\cal B} (\rho, \phi)}{\partial \rho} \,
 d \rho \wedge d \phi.  \label{Eq:F_electric_iii}
\end{eqnarray}
Here, we have introduced a new auxiliary function $\hat{\cal B} (\rho, \phi)$ by 
\begin{equation}
 \hat{\cal B} (\rho, \phi) \ \equiv \ B ( G(\rho) \,g (\phi)).  \label{Bhat}
\end{equation}
It satisfies the following equation
\begin{equation}
 \frac{\partial \hat{\cal B} (\rho, \phi)}{\partial \rho} \ = \ \left. 
 G^\prime (\rho) \,g (\phi) \,\dot{B} (t) \,\right\vert_{t = G(\rho) \,g (\phi)} .
\end{equation}
Inserting the above relation into Eq.(\ref{Eq:F_electric_iii}), and 
integrating it over $\rho$ from $0$ to $R$, we obtain
\begin{eqnarray}
 - \,\int_0^R \,d \rho \,\frac{\rho^2}{2} \,
 \frac{\partial \hat{\cal B} (\rho, \phi)}{\partial \rho}
 \ = \ - \,\frac{R^2}{2} \,\hat{\cal B} (R, \phi) \ + \ 
 \int_0^R \,d \rho \,\rho \,\hat{\cal B} (\rho, \phi). \hspace{3mm}
\end{eqnarray}
In this way, it turns out that the electric field contribution to the surface integral 
from the region (iii) is given by
\begin{eqnarray}
 \left. \iint_{region \ \rm{(iii)}} \,\frac{1}{2} \,F_{\mu \nu} \,d x^\mu \wedge d x^\nu
 \,\right\vert_{electric} 
 \ =  \ \int_{\phi_f}^{\phi_i + 2 \,\pi} d \phi \,\left[
 - \,\frac{R^2}{2} \,\hat{\cal B} (R, \phi) + \int_0^R \,d \rho \,\rho \,
 \hat{\cal B} (\rho, \phi) \right] .  
\end{eqnarray}

\vspace{2mm}
\noindent
(iv) contribution from the region $\phi_f \leq \phi \leq \phi_i + 2 \,\pi, \ R \leq \rho < \rho_0$

\vspace{2mm}
Carrying out a similar manipulation, we can show that the electric field contribution 
to the surface integral from the region (iv) is obtained as
\begin{eqnarray}
 \left. \iint_{region \ \rm{(iv)}} \,\frac{1}{2} \,F_{\mu \nu} \,d x^\mu \wedge d x^\nu
 \,\right\vert_{electric}
 \ = \ \int_{\phi_f}^{\phi_i + 2 \,\pi} d \phi \,\left[
  - \,\frac{R^2}{2} \,\hat{\cal B} (\rho_0, \phi) \ + \ 
 \frac{R^2}{2} \,\hat{\cal B} (R, \phi) \right] . 
\end{eqnarray}

\vspace{3mm}
\noindent
Now combining the contributions from the regions (i) and (ii), we have
\begin{eqnarray}
 &\,& \left. \iint_{region \ \rm{(i)} \, + \, \rm{(ii)}} \,\frac{1}{2} \,F_{\mu \nu} \,
 d x^\mu \wedge d x^\nu
 \,\right\vert_{electric} \nonumber \\
 &\,& = \int_{\phi_i}^{\phi_f} \! d \phi \,\left[
 - \,\frac{R^2}{2} \,\tilde{\cal B} (R, \phi) \ + \ \int_0^R \,d \rho \,\rho \,
 \tilde{\cal B} (\rho, \phi) \right]
 \ + \ \int_{\phi_i}^{\phi_f} \! d \phi \,\left[
  - \,\frac{R^2}{2} \,\tilde{\cal B} (\rho_0, \phi) \ + \ 
 \frac{R^2}{2} \,\tilde{\cal B} (R, \phi) \right] \hspace{1mm} \nonumber \\ 
 &\,& = \int_{\phi_i}^{\phi_f} \! d \phi \,\left[
 - \,\frac{R^2}{2} \,\tilde{\cal B} (\rho_0, \phi) \ + \ \int_0^R \,d \rho \,\rho \,
 \tilde{\cal B} (\rho, \phi) \right] . \hspace{4mm}
\end{eqnarray}
Similarly, the contribution from the regions (iii) and (iv) are given by
\begin{eqnarray}
 \left. \iint_{region \ \rm{(iii)} \, + \, \rm{(iv)}} \,\frac{1}{2} \,F_{\mu \nu} \,
 d x^\mu \wedge d x^\nu
 \,\right\vert_{electric} 
 \ = \ \int_{\phi_f}^{\phi_i + 2 \,\pi} \!\! d \phi \,\left[
 - \,\frac{R^2}{2} \,\hat{\cal B} (\rho_0, \phi)  +  \int_0^R \,d \rho \,\rho \,
 \hat{\cal B} (\rho, \phi) \right] . 
\end{eqnarray}
Summing up the contributions from the region (i) + (ii) and the region (iii) + (iv),
the net electric field contribution from the whole surface $S$ is given by
\begin{eqnarray}
 \left. \iint_{S \,(C_1 - C_2)} \frac{1}{2} \,F_{\mu \nu} \,d x^\mu \wedge d x^\nu \,
 \right\vert_{electric} 
 \!\!\!&=& \int_{\phi_i}^{\phi_f} d \phi \left[ - \,\frac{R^2}{2} B (f(\phi)) +  
 \int_0^R \!d \rho \,\rho \,B (F(\rho) \,f (\phi)) \right] \hspace{6mm} \nonumber \\
 \!\!\!&+&  \int_{\phi_f}^{\phi_i + 2 \,\pi} d \phi 
 \left[ - \,\frac{R^2}{2} B (g(\phi))   +   
 \int_0^R \!\!\!d \rho \,\rho \,B (G(\rho) \,g (\phi)) \right]. \hspace{6mm} 
\end{eqnarray}
Here, made has been use of the relations 
$\tilde{\cal B} (\rho, \phi) = B (F (\rho) \,f (\phi)), \,
\hat{\cal B} (\rho, \phi) = B (G (\rho) \,g (\phi))$
together with the constraint conditions $F (\rho_0) = 1$ and $G (\rho_0) = 1$.

\subsection{Magnetic field contribution}

\vspace{1mm}
Next, let us evaluate the magnetic field contribution to the surface integral.
Using the relations
\begin{equation}
 F_{yz} = F_{zx}  =  0, \ \  \ 
 F_{xy}  =  \left\{ \begin{array}{cl}
 - \, B(t) \ & \ (0 \leq \rho < R), \\
 0 \  & \  (\rho > R), \\
 \end{array} \right.
\end{equation}
we have
\begin{eqnarray}
 \left. \frac{1}{2} \,F_{\mu \nu} \,d x^\mu \wedge d x^\nu \,\right\vert_{magnetic}
 &=& F_{xy} \,\,d x \wedge d y \nonumber \\
 &=& F_{xy} \left( \cos \phi \,d \rho - \rho \,\sin \phi \,d \phi \right) \wedge
 \left( \sin \phi \,d \rho + \rho \,\cos \phi \,d \phi \right)  \hspace{4mm} \nonumber \\
 &=&  - \,B (t) \,\rho \, \,d \rho \wedge d \phi ,
\end{eqnarray}
in the region $\rho < R$. Combining it with the observation that there is no
magnetic field in the region $\rho > R$, we therefore find that the magnetic field
contribution to the surface integral is given by
\begin{eqnarray}
 \left. \iint \frac{1}{2} \,F_{\mu \nu} \,d x^\mu \wedge d x^\nu \,
 \right\vert_{magnetic} 
 &=&  - \,\int_0^R \rho \,d \rho \,\int_{\phi_i}^{\phi_f} d \phi \,
 B (F (\rho) \,f (\phi)) \nonumber \\
 &\,&   - \int_0^R \rho \,d \rho \,
 \int_{\phi_f}^{\phi_i + 2 \,\pi} d \phi \,B (G (\rho) \, g (\phi)). \hspace{8mm}
\end{eqnarray}

\subsection{ Total contribution to the area integral}

Now we are ready to sum up the contributions to the area integral from the
electric part and the magnetic part.
\begin{eqnarray}
 &\,& \left. \iint_{S \,(C_1 - C_2)} \frac{1}{2} \,F_{\mu \nu} \,d x^\mu \wedge d x^\nu \,
 \right\vert_{electric}
 \ + \  
  \left. \iint_{S \,(C_1 - C_2)} \frac{1}{2} \,F_{\mu \nu} \,d x^\mu \wedge d x^\nu \,
 \right\vert_{magnetic} \hspace{5mm} \nonumber \\
 &\,& = \int_{\phi_i}^{\phi_f} d \phi \,\left\{ - \,\frac{R^2}{2} \,B (f (\phi)) 
 \ + \ 
 \int_0^R \,d \rho \,\rho \,\int_{\phi_i}^{\phi_f} \,B (F(\rho) \,f (\phi)) \right\} 
 \nonumber \\
 &\,& + \int_{\phi_f}^{\phi_i + 2 \,\pi} d \phi \,\left\{ - \,\frac{R^2}{2} \,B (g (\phi)) 
 \ + \  
 \int_0^R \,\rho \,d \rho \,\int_{\phi_f}^{\phi_i + 2 \,\pi} d \phi \,
 B ( G(\rho) \,g (\phi)) \right\} \nonumber \\
 &\,& - \int_0^R \,d \rho \,\rho \,\int_{\phi_i}^{\phi_f} \,B (F (\rho) \,f (\phi)) 
 \ - \  \int_0^R \,\rho \,d \rho \,\int_{\phi_f}^{\phi_i + 2 \,\pi} \!\!\! d \phi \,
 B (G (\rho) \,g (\phi)) .
\end{eqnarray}
Remarkably, we find that the parts containing the integration over $\rho$ from 
$0$ to $R$ cancel among the electric and magnetic contributions.  
As a consequence, the net contribution to the area integral from the electric and 
magnetic field are given as
\begin{eqnarray}
 \left. \iint_{S \,(C_1 - C_2)} \frac{1}{2} \,F_{\mu \nu} \,d x^\mu \wedge d x^nu \,
 \right\vert_{electric + magnetic}
 \!\!\! = \   
 - \,\frac{R^2}{2} \! \int_{\phi_i}^{\phi_f} \! d \phi \,B (f (\phi))  -  
 \frac{R^2}{2} \! \int_{\phi_f}^{\phi_i + 2 \,\pi} \!\!\! d \phi \,B (g (\phi)) . 
 \hspace{2mm} \label{Eq:area_integ}
\end{eqnarray}

\subsection{4-dimensional Stokes theorem}

Using the results obtained so far, we can verify that the 4-dimensional
Stokes theorem certainly holds. To confirm it, we need to evaluate the
closed line-integral $\oint_{C_1 - C_2} A_\mu \,d x^\mu$ by using the
present parametrization of the space-time paths $C_1$ and $- \,C_2$.
Using
\begin{equation}
 \left. \bm{A} \,\right\vert_{\rho = \rho_0} \ = \ \frac{R^2}{2 \,\rho_0} \,B (t) \,\,
 \left( - \,\sin \phi, \cos \phi, 0 \right) ,
\end{equation}
we obtain
\begin{eqnarray}
 \left.  A_\mu \,d x^\mu \,\right\vert_{\rho = \rho_0} 
 &=&
 - \,A_x \,d x \ - \ A_y \,d y \nonumber \\
 &=& \frac{R^2}{2 \,\rho_0} \, B(t) \,\sin \phi \,
 \left( \cos \phi \,d \rho \ - \ \rho \,\sin \phi \,d \phi \right) \nonumber \\
 &-& 
 \frac{R^2}{2 \,\rho_0} \,B(t) \,\cos \phi \,
 \left( \sin \phi \,d \rho \ + \ \rho \,\cos \phi \,d \phi \right) \nonumber \\
 &=&  - \,\frac{R^2}{2} \, B(t) \, d \phi 
 \ =\  \left\{ \begin{array}{cl}
 - \,\frac{R^2}{2} \,B (f(\phi)) \,d \phi \ \ & \ \ (\phi_i \leq \phi \leq \phi_f) , \\
 - \,\frac{R^2}{2} \,B (g(\phi)) \,d \phi \ \ & \ \ (\phi_f \leq \phi \leq \phi_i + 2 \,\pi) . \\
 \end{array} \right.  \hspace{5mm}
\end{eqnarray}
It gives
\begin{eqnarray}
 \oint_{C_1 - C_2} A_\mu \,d x^\mu &=& 
 - \,\frac{R^2}{2} \,\int_{\phi_i}^{\phi_f} \,d \phi \,B (f (\phi)) 
 \ - \ 
 \frac{R^2}{2} \,\int_{\phi_f}^{\phi_i + 2 \,\pi} \!\!\! d \phi \,B (g (\phi)).
 \hspace{6mm}
\end{eqnarray}
This precisely coincides with the previously evaluated area integral 
(\ref{Eq:area_integ}) of the field-strength tensor, which
shows that the 4-dimensional Stokes theorem in the form
\begin{eqnarray}
 \left. \iint_{S \,(C_1 - C_2)} \frac{1}{2} \,F_{\mu \nu} \,d x^\mu \wedge d x^\nu \,
 \right\vert_{electric + magnetic} 
 \ = \   
 \oint_{C_1 -C_2} \,A_\mu \,d x^\mu ,
\end{eqnarray}
certainly holds also in our particular setting. 

Note, however, that our explicit derivation 
of this general mathematical identity in the special setting of the 
infinitely-long solenoid with the time-varying magnetic flux leads us
to deeper understanding about the physical implication of 
this mathematical theorem.
Note that the l.h.s of the above identity is given as a 2-dimensional area
integral, while the r.h.s. is given as a 1-dimensional closed-line integral.
This already implies that, for the identity to hold, the 2-dimensional
surface integral part of the l.h.s. must be cancelled out in some way. 
In fact, we have shown that such a cancellation occurs between the 
contributions from the induced electric field and the magnetic field.
On might then tempted to think that this observation might support the 
cancellation argument by Singleton et al. However, one must remember the
fact that, in the analyses by Singleton et al., this cancellation is claimed to happen 
only between the {\it time-dependent part} of the magnetic field and the 
induced electric field. 
Accordingly, they claim that only the {\it time-independent part} of the 
magnetic field contribute to the Aharonov-Bohm phase shift even in the situation
where the magnetic field varies time-dependently, which amounts to claiming
the absence of the time-dependent AB-effect.
This conclusion is not correct, however.
Although the cancellation between the contributions of the magnetic field 
and the induced electric field to the surface integral certainly happens,
what remains after this cancellation is the closed space-time line integral of 
the 4-vector potential. 
As shown in the subsequent sections, this closed line integral is most
likely to generate non-trivial time-dependence to observable AB-phase shift.

\section{\label{sec5}What can we say about the time-dependent AB-effect, after all ? }

Through the analyses in the previous two sections, we have confirmed that
the AB-phase shift can generally be expressed as space-time line-integrals 
in either of the two equivalent forms as follows,
\begin{eqnarray}
 \phi_{AB}  &=& \ - \,e \, \oint_{C_1 - C_2} A_\mu \,d x^\mu
 \ = \ - \,e \,\left\{ \int_{C_1} A_\mu \,d x^\mu \ -  \ \int_{C_2} A_\mu \,d x^\mu \right\},
 \hspace{6mm}
\end{eqnarray}
even in the case where the magnetic flux penetrating the solenoid changes
time-dependently as $\bm{B} (\bm{x}, t) = B (t) \,\bm{e}_z$.
For convenience, here we continue our analysis with the use of the 2nd form of
the above expression.
Supposing that path $C_1$ is taken to be a circular arc around the coordinate origin
with radius $\rho \,( \,> R)$, which begins from the azimuthal angle $\phi = \phi_i$ 
and ends at $\phi = \phi_f$, the space-time points on the path $C_1$ are 
parametrized as
\begin{equation}
 t (\phi) = f (\phi), \ \ \phi_i \ \leq \ \phi \ \leq \ \phi_f  .
\end{equation}
with $f (\phi_i) = t_i, \, f (\phi_f) =  t_f$.
Similarly, supposing that path $C_2$ is taken to be a semi-circle with radius 
$\rho \,(\, > R)$, which begins from $\phi = \phi_i$ and ends at 
$\phi = \phi_f - 2 \,\pi$, the space-time points on the path $C_2$ are parametrized as
\begin{equation}
 t (\phi) = g (\phi), \ \ \phi_f - 2 \,\pi \ \leq \ \phi \ \leq \ \phi_i , 
\end{equation}
with $g (\phi_i) = t_i, \, g (\phi_f - 2 \,\pi )  =  t_f$.
Then, following a similar analysis in the previous section, we can show that
\begin{eqnarray}
 \int_{C_1} A_\mu \,d x^\mu &=& - \,\frac{R^2}{2} \,
 \int_{\phi_i}^{\phi_f} d \phi \,B (f (\phi)), \\ 
 \int_{C_2} A_\mu \,d x^\mu &=& - \,\frac{R^2}{2} \,
 \int_{\phi_i}^{\phi_f - 2 \,\pi} d \phi \,B (g (\phi)).  \hspace{6mm}
 \label{Eq:AB-phase_final}
\end{eqnarray}
One should clearly understand the precise meaning of the above formulas, which  
unambigously shows that, on the space-time paths $C_1$ and $C_2$, the variable 
$t$ and $\phi$ cannot be treated as independent integration variables. 
As we shall see below, this turns out to be vitally important in the proper 
treatment of the time-dependent AB-effect.

\vspace{2mm}
Although it may not be so easy to recognize, highly nontrivial 
effects of induced electric field is mixed in with the above expressions of the 
line-integral of the 4-vector potential.
In fact, in the case where the magnetic flux inside the solenoid is time-dependent,
non-zero electric field is generated even outside the solenoid,
and this induced electric field outside the solenoid is thought to affect 
the motion of the charged particle.
As a warm-up to handling this fairly subtle nature of the problem,
we think it instructive to begin the analysis with much simpler case of the 
familiar time-independent AB-effect.
Since no electromagnetic force acts on the charged particle moving outside the 
solenoid in this case, one may simply assume that the charged particle makes 
a rotational motion around the origin with a {\it constant} angular velocity $\omega$. 
The simplest parametrization of the function $f (\phi)$ on the path $C_1$ is 
therefore given by (Here, we set the initial time $t_i$ to be zero, for simplicity.) 
\begin{equation}
 t \ \equiv \ f (\phi)  \ = \ \frac{\phi}{\omega} \ \ \ \mbox{or} \ \ \ 
 \phi \ = \ \omega \,t,
\end{equation}
with $f (\phi_i) = t_i \equiv 0, \, \phi_i = 0$ and  $f (\phi_f) =  t_f , \ \phi_f = \pi$,
%
%
%
which dictates that $t_f = \pi / \omega$. Since the magnetic flux here is
time-independent, i.e. $B(t) = B_0$, we obtain
\begin{eqnarray}
 \int_{\phi_i}^{\phi_f} d \phi \,B (f (\phi)) &=& \int_0^{t_f} \,
 \omega \,d t \,\,B_0
 \ = \ \omega \,B_0 \,\int_0^{\pi / \omega} \,d t 
 \ = \ \pi \,B_0, \hspace{8mm}
\end{eqnarray}
which is independent of $\omega$.
Similarly, the simplest parametrization of $g (\phi)$ on $C_2$ is 
given by
\begin{equation}
 t \ \equiv \ g (\phi)  = - \,\frac{\phi}{\omega} \ \ \ \mbox{or} \ \ \ 
 \phi = - \,\omega \,t,
\end{equation}
with
\begin{equation}
 g (\phi_i) = t_i = 0, \ \phi_i = 0, 
\end{equation}
and
\begin{equation} 
 g (\phi_f - 2 \,\pi)  =  t_f , \ \phi_f  -  2 \,\pi = - \,\pi, 
\end{equation}
which also means that $t_f  = \pi / \omega$.
Then, for $B (t) = B_0$, we have
\begin{eqnarray}
 \int_{\phi_i}^{\phi_f - 2 \,\pi} \!\!\! d \phi \,B (g (\phi)) &=& 
 \int_0^{- \,\pi} \! d \phi \,B_0
 \ =\ 
 B_0 \,\int_0^{t_f} ( \,- \,\omega) \, d t = - \,\pi \,B_0. \hspace{8mm}
\end{eqnarray}
Using the above results, we obtain
\begin{eqnarray}
 \phi_{AB} &=& - \,e \,\left\{ - \,\frac{R^2}{2} \,\pi \,B_0  \ - \   
 \left( + \,\frac{R^2}{2} \,\pi \,B_0 \right) \right\} 
 \ = \ e \,\pi \,R^2 \,B_0
 \ = \ e \,\,\Phi_0.
\end{eqnarray}
This precisely reproduces the familiar AB-phase shift in the case of 
time-independent magnetic flux.

\vspace{2mm}
After the above warming-up exercise, we turn to the time-varying magnetic
flux case which is of our real concern.
Main difficulty in this case is that the angular velocity of the charged
particle cannot be regarded as a constant anymore. This is because the 
angular velocity of the charge particle is changed by the force due to 
the induced electric field, which exerts along the circumferential direction.
This makes the theoretical treatment of the time-dependent AB-effect
quite complicated.  At present, we do not have  
completely satisfactory solution to this difficult problem.
In the following, we therefore consider two simple cases in which we can
treat it at least within a reasonable approximation.
The first is the case in which the change of the time-dependent magnetic 
flux is very slow so that the change of the angular velocity of the charge
particle is negligibly small within the time interval between $t_i = 0$ and $t_f$.
Another is the case  in which the magnetic flux is rapidly and sinusoidally
oscillating. If the time-period of this sinusoidal oscillation is much
shorter than the time interval between $t_i = 0$ and $t_f$,
the charged particle feels oscillating Lorentz force along the path
due to the induced electric field. This implies that, on the average, 
the angular velocity of the charged particle can approximately be treated as 
a time-independent constant. Under these circumstances, we may be
able to proceed as follows.  
That is, the only difference from the above-explained treatment with
the time-independent magnetic flux is that the magnetic field $B (t)$
is time-dependent.
Then, the 1st integral on the r.h.s. of Eq.(\ref{Eq:AB-phase_final}) 
can be evaluated as
\begin{eqnarray}
 &\,& \int_{\phi_i}^{\phi_f} \,d \phi \,B (f (\phi)) =  
 \int_0^{\pi / \omega} \,\omega \,\,d t \,\, B (t)
 \ = \ 
 \omega \,\int_0^{t_f} \,d t \,B (t)
 \ = \ 
 \omega \,\bar{B} (t_f) \ = \ \pi \,\,\frac{\bar{B}(t_f)}{t_f}, \hspace{8mm}
\end{eqnarray}
where we have introduced an auxiliary function $\bar{B} (t)$ by
\begin{equation}
 \bar{B} (t) \ \equiv \ \int_0^t \,d t^\prime \,B (t^\prime),
\end{equation}
and also used the relation $t_f = \pi / \omega$. 

\vspace{2mm}
Similarly, the 2nd integral on the r.h.s. of Eq.(\ref{Eq:AB-phase_final}) 
are evaluated as follows. 
As a  convenient parametrization of $C_2$ and $g (\phi)$, we choose
\begin{equation}
 t \ = \ g (\phi) \ = \ - \,\frac{\phi}{\omega}, \ \ \ \mbox{or} \ \ \ 
 \phi \ = \ - \,\omega \,t,
\end{equation} 
with $g (\phi_i) = t_i = 0, \, \phi_i = 0$ and 
$g (\phi_f - 2 \,\pi) = t_f , \, \phi_f  - 2 \,\pi = - \,\pi$,
%
 %
 which dictates that $t_f = \pi / \omega$. We thus obtain
 \begin{eqnarray}
  \int_{C_2} \, d \phi \,B (g(\phi))  =  
  \int_0^{t_f} \,\left( - \,\omega \,d t \right) \,B (t)
   \ = \   - \,\omega \,\bar{B} (t_f) = - \,\pi \,\,\frac{\bar{B} (t_f)}{t_f}.
  \hspace{8mm}
 \end{eqnarray}
 In this way, we eventually get
 \begin{eqnarray}
  \phi_{AB} \ = \ e \,\,\frac{R^2}{2} \,\left\{ \,\pi \,\frac{\bar{B} (t_f)}{t_f} 
  \ - \ (\, - \,\pi ) \,\frac{\bar{B} (t_f)}{t_f} \right\}
  \ = \ 
  e \,\pi \,R^2 \,\,\frac{\bar{B} (t_f)}{t_f} .
 \end{eqnarray}
 Using the total magnetic flux $\Phi (t) \equiv \pi \,R^2 \,B (t)$ inside
 the solenoid, this answer can also be expressed in the form
 \begin{equation}
  \phi_{AB} \ = \ e \,\,\frac{\bar{\Phi} (t_f)}{t_f},
 \end{equation}
with the definition
 \begin{equation}
  \bar{\Phi} (t) \ \equiv \ \int_0^t \,d t^\prime \,\Phi (t^\prime) .
 \end{equation}
As a minimal consistency check, let us consider the limit, in which the magnetic
flux is time-independent, i.e. $\Phi (t) = \Phi_0$. In this case, we get
\begin{equation}
 \bar{\Phi} (t_f) \ = \ \int_0^{t_f} \,\Phi_0 \, d t \ = \ \Phi_0 \,t_f ,
\end{equation}
which legitimately reproduces the familiar expression of the time-independent
AB-phase shift, since
\begin{equation}
 \phi_{AB} \ = \ e \, \, \frac{\Phi_0 \,t_f}{t_f} \ = \ e \,\,\Phi_0 .
\end{equation}
Practically more interesting is the case where the magnetic flux is rapidly
and sinusoidally oscillating as considered in the paper by 
Lee et al. \cite{LYGC1992}. 
(It can in principle be generated by oscillating electric current
flowing on the surface of a very long solenoid.) For example,
suppose that $\Phi (t)$ is given by
\begin{equation}
 \Phi (t) \ = \ \Phi_0 \,\cos \Omega \,t.
\end{equation}
In this case, we obtain
\begin{equation}
 \bar{\Phi} (t_f) \ = \ \Phi_0 \,\,\frac{\sin \Omega \,t_f}{\Omega},
\end{equation}
which in turn gives
\begin{equation}
 \phi_{AB} \ = \ e \,\,\Phi_0 \,\frac{\sin \Omega \,t_f}{\Omega \,t_f} .
\end{equation}
This clearly shows that, contrary to the claim by Singleton et al.,
the time-dependent AB-effect is most likely to exist as claimed by many 
researchers \cite{LYGC1992,ADC2000, GNS2011,JZWLD2017,CM2019},
although there are diversities in their theoretical treatments as well as
predictions.

\section{\label{sec6}Concluding remarks}

We have investigated anew the controversy over the time-dependent AB effect.
The central objective of our study is to verify the validity of the claim 
that the effect of time-dependent magnetic vector potential to the AB-phase shift
is precisely cancelled by the effect of induced electric field so that
there is no time-dependent AB-effect. For that purpose, we go back to basics of 
quantum mechanics and move forward step by step.
First, we carefully analyze the Schr\"{o}dinger equation of the electron under the
influence of the time-dependent magnetic field confined inside the solenoid.
We have then confirmed that this analysis provides a firm theoretical foundation
to the 1st covariant expression of the time-dependent AB-phase shift, 
which is given as a closed line integral of the vector potential 
in the 4-dimensional Minkowski space.
Next, since the cancellation argument by Singleton et al. is based on the 4-dimensional 
Stokes theorem, which states that the closed line integral of the 4-vector potential
is also expressed as a surface integral of the electromagnetic field strength tensor, 
we carefully re-investigated the derivation of the 4-dimensional Stokes theorem
in line with our specific setting of infinitely-long solenoid with time-dependent
magnetic flux. Through the explicit calculation of the surface integral of the field 
strength tensor, we have certainly confirmed that a sort of cancellation 
occurs between the magnetic and induced electric field contributions
to the surface integral.  However, what remains after this cancellation is
the closed line-integral of the 4-vector potential in the Minkowski space,
which is nothing but what is meant by the 4-dimensional Stokes theorem.
We have thus confirmed once again that the time-dependent AB-phase shift
is given as the closed space-time line integral of the 4-vector potential. 

\vspace{1mm}
Unfortunately, we realize that exact or completely satisfactory evaluation 
of this closed line integral is a highly nontrivial task.
In some previous literature \cite{JZWLD2017}, the evaluation of this closed line integral is
carried out by following the logic from Eq.(\ref{Eq:Stokes3_1}) to Eq.(\ref{Eq:Stokes3_3}),
especially relying upon Eq.(\ref{Eq:Stokes3_2}) in the present paper, which amounts 
to naively assuming that the Stokes theorem in the 3-dimensional form can be applied here.
This manipulation is justified if the vector potential $\bm{A}$ contained in 
Eq.(\ref{Eq:Stokes3_2}) is time-independent, but we found that it is not justified 
in the case the vector potential $\bm{A}$ is time-dependent.
Consequently, the prediction for the time-dependent AB-phase shift 
given by Eq.(\ref{Eq:Stokes3_3}) is not correct as was suspected there.
The point is that $C_1$ and $C_2$ appearing in Eq.(\ref{Eq:Stokes3_1}) is a space-time
paths in the Minkowski space. This means that the spatial variable $\bm{x}$ and
time variable $t$ cannot be regarded as totally independent integration variables
within the line integrals $\int_{C_1} \,A_\mu (x) \,d x^\mu$ and
$\int_{C_2} \,A_\mu (x) \,d x^\mu$.
As discussed in sect \ref{sec5}, what makes evaluation of this integral difficult is the 
presence of the induced electric field, which exists even outside of the solenoid. 
This induced electric field exerts Lorenz force on the charged particle, thereby 
changing its velocity and/or angular velocity.   
There are two cases in which at least approximate evaluation of the space-time
line-integral is possible. The first is the case in which the time-dependence
of the magnetic field inside the solenoid is weak enough so that the 
effect of the induced electric field outside the solenoid is very small.
In such a circumstance, the change of the electrons velocity and/or the angular 
velocity during the time interval in which the charged particle passes around
the solenoid is negligible. Practically more interesting would be the 
case in which the magnetic flux inside the solenoid is rapidly and sinusoidally 
oscillating.
If the time-period of this sinusoidal oscillation is much shorter than the
time scale in which the charged particle passes around the solenoid, the charged 
particle would feel oscillating or alternating Lorenz force along the path.
Then, on the average, the velocity or angular velocity of the charged particle
may approximately be treated as a time-independent constant.
Under these circumstances, we can give explicit theoretical prediction, 
which at the least definitely confirms the existence of the time-dependent AB-effect. 
A real remaining question is how to experimentally confirm the existence
of the time-dependent AB-effect. We conjecture that the simplest but most 
likely experiments would be the following. 
Suppose that we change the magnetic field inside the solenoid rapidly and 
sinusoidally, while continuously providing the charged particle beam 
directed to the double-slit panel. If our theoretical prediction is meaningful, 
the rapid change or oscillation of the interference pattern on the screen
is expected to be seen. Undoubtedly, this kind of experiment would be feasible 
even for more realistic setting by using a troidal coil as adopted by Tonomura et al.'s 
experiment \cite{Tonomura1986}.

\section*{Acknowledgement}
The author of the paper would like to thank Takahiro Kubota
for several pieces of advice and enlightening discussion especially
for the derivation of the 4-dimensional Stokes theorem.
\

%

\appendix

\section{On the path-independence of the gauge function $\Lambda (x)$
defined by Eq.(\ref{Eq:gauge_function_Lambda}).}

Consider two space-time paths $C_1$ and $C_2$, both of which start
from a fixed space-time point $x^\mu_0$ and end at the point $x^\mu$.
The gauge functions $\Lambda (x ; C_1)$ and $\Lambda (x ; C_2)$ are
respectively defined by the following line integrals :
\begin{eqnarray}
  \Lambda (x ; C_1) &\equiv& - \,\int_{C_1} \,A_\mu (x^\prime) \,d x^{\prime \mu}, \\
  \Lambda (x ; C_2) &\equiv& - \,\int_{C_2} \,A_\mu (x^\prime) \,d x^{\prime \mu}.
  \hspace{6mm}
\end{eqnarray}
The difference between these two functions is given by
\begin{eqnarray}
 \Lambda (x ; C_1) \ - \ \Lambda (x ; C_2)
 \ =\ 
 \oint_{C_1 - C_2} \,A_\mu (x^\prime) \,d x^{\prime \mu}
 \ = \  
 \iint_{S \,(C_1 - C_2)} \frac{1}{2} \,F_{\mu \nu} (x^\prime) \,
 d {x^{\prime \mu}} \wedge d {x^{\prime \nu}}, \hspace{4mm}
\end{eqnarray}
Here, at the last step, we have used the 4-dimensional Stokes theorem.

\begin{figure}[ht]
\begin{center}
\includegraphics[width=6.5cm]{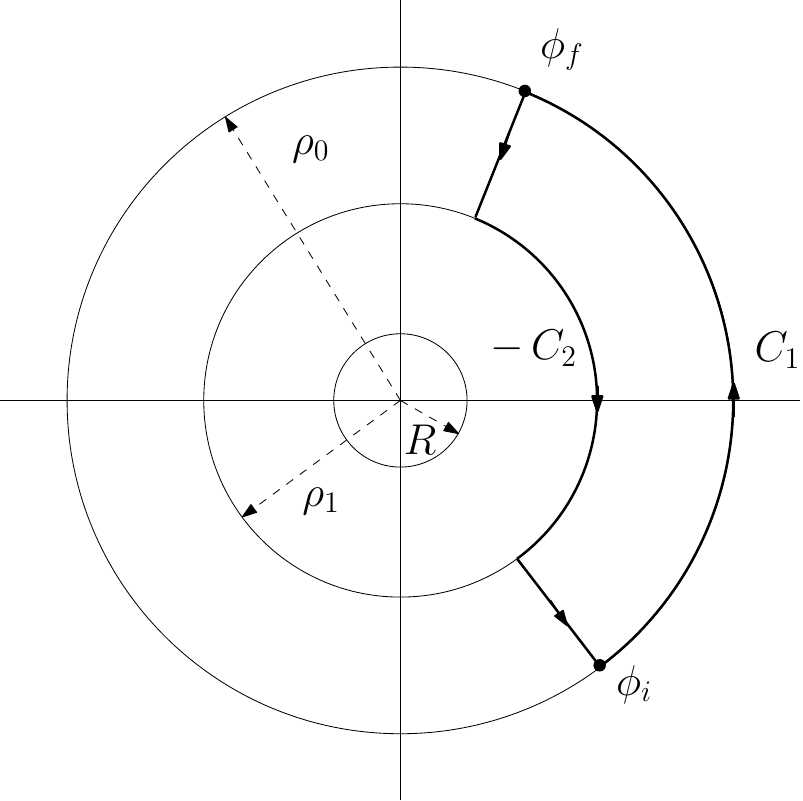}
\caption{A closed loop $C_1 - C_2$ in the 4-dimensional Minkowski space,
which does not encircle the solenoid.}
\label{Fig:OuterLoop}
\end{center}
\end{figure}

\vspace{2mm}
Now we consider the case in which the closed space-time path $C_1 - C_2$
does not encircle the solenoid. For simplicity, we assume that this closed 
path is given in the form depicted in Fig.\ref{Fig:OuterLoop}. 
The path $C_1$ is a circular arc with
radius $\rho_0 \,(> R)$ with the initial and final azimuthal angles
respectively given by $\phi_i$ and $\phi_f$. 
On the other hand, the path $- \,C_2$
is given as a sum of three paths. The 1st is the straight-line path
that starts from $\rho = \rho_0$ and ends at $\rho = \rho_1 \,
( R < \rho_1 < \rho_0)$ while keeping the angle $\phi$ to $\phi_f$.
The 2nd is the circular arc with radius $\rho_1$  with the
initial and final azimuthal angles respectively given by 
$\phi_f$ and $\phi_i$.
The 3rd is the straight-line path that starts from $\rho = \rho_1$
and ends at $\rho = \rho_0$ while keeping the angle $\phi$ to $\phi_i$.

\vspace{2mm}

We first evaluate the contribution of the magnetic field to
the surface integral.
Proceeding in the same way as in sect.IV, we generally have
\begin{eqnarray}
 \left. \frac{1}{2} \,F_{\mu \nu} \,\,d x^\mu \wedge d x^\nu \,\right\vert_{magnetic}
 \ = \ F_{xy} \,d x^\mu \wedge d x^\nu
 \ = \  
 F_{xy} \,\,\rho \,d \rho \wedge d \phi . \hspace{5mm}
\end{eqnarray}
Note however that, in the region surrounded by the space-time $C_1 - C_2$, 
it holds that
\begin{equation}
 F_{xy} \ = \ 0 \ \ \ \mbox{for} \ \ \ R < \rho_1 < \rho < \rho_0 .
\end{equation}
This means that
\begin{equation}
  \left. \frac{1}{2} \,\,F_{\mu \nu} \,d x^\mu \wedge d x^\nu \,\right\vert_{magnetic}
  \ = \ 0.
\end{equation}
It seems to be consistent with the intuition, since
the magnetic flux of the solenoid is not included inside the
closed space-time curve $C_1 - C_2$.

\vspace{2mm}
A little more nontrivial is the contribution of the induced electric field, which
extends even outside the solenoid.
In the expression
\begin{eqnarray}
 \left. \frac{1}{2} \,F_{\mu \nu} \,\,d x^\mu \wedge d x^\nu \,\right\vert_{electic}
 \ = \ F_{0x} \,\,dt \wedge dx \ + \ F_{0y} \,\,dt \wedge dy \ + \ F_{0z} \,\,dt \wedge dz,
 \hspace{2mm}
\end{eqnarray}
we use the fact that, for $R < \rho_1 < \rho < \rho_0$, it holds that
\begin{equation}
 F_{0x} = \frac{R^2}{2} \dot{B} (t) \,\frac{1}{\rho} \sin \phi, \ 
 F_{0y} = - \,\frac{R^2}{2} \dot{B} (t) \,\frac{1}{\rho} \cos \phi, \  
 F_{0z} =  0.
\end{equation}
We thus find that
\begin{equation}
 \left. \frac{1}{2} \,F_{\mu \nu} \,d x^\mu \wedge d x^\nu \,\right \vert_{electic}
 \ = \ - \,\frac{R^2}{2} \,\frac{\partial \tilde{\cal B} (\rho, \phi)}{\partial \rho} \,
 d \rho \wedge d \phi. 
\end{equation}
Here, $\tilde{\cal B} (\rho, \phi)$ is an auxiliary function defined in (\ref{Btilde}).
In this way, the contribution of the induced electric field to the surface
integral turns out to be
\begin{eqnarray}
 \left. \frac{1}{2} \,F_{\mu \nu} \,d x^\mu \wedge d x^\nu \,\right \vert_{electic}
 &=& \int_{\phi_i}^{\phi_f} d \phi \int_{\rho_1}^{\rho_0} d \rho \,\left( 
 - \,\frac{R^2}{2} \,\frac{\partial \tilde{\cal B} (\rho, \phi)}{\partial \rho} \right)
 \nonumber \\
 &=& \int_{\phi_i}^{\phi_f} \, d \phi \,
 \left( - \,\frac{R^2}{2} \,\tilde{\cal B} (\rho_0 ,\phi)
 + \frac{R^2}{2} \,\tilde{\cal B} (\rho_1, \phi) \right) . \hspace{5mm}
\end{eqnarray}
Incidentally, from the requirement that the point
$(t (\rho_0, \phi), x (\rho_0, \phi), y (\rho_0, \phi), z (\rho_0, \phi)$ lie on the
curve $C_1$, we have $F (\rho_0) = 1$.
Similarly, from the requirement that the point
$(t (\rho_1, \phi), x (\rho_1, \phi), y (\rho_1, \phi), z (\rho_1, \phi)$ lie on the
curve $- \,C_2$, we have $F (\rho_1)  =  1$.
%
%
This dictates that
\begin{eqnarray}
 \tilde{\cal B} (\rho_0, \phi) &=& B (F(\rho_0) \,f (\phi)) \ = \ B (f(\phi)), \\
 \tilde{\cal B} (\rho_1, \phi) &=& B (F(\rho_1) \,f (\phi)) \ = \ B (f(\phi)). \hspace{8mm}
\end{eqnarray}
As a consequence, we eventually get
\begin{eqnarray}
 \iint_{S \,(C_1 - C_2)} 
 \left. \frac{1}{2} \,F_{\mu \nu} \,d x^\mu \wedge d x^\nu \,\right \vert_{electic}
 \ = \  \int_{\phi_i}^{\phi_f} \left(\, - \,\frac{R^2}{2} \, B(f(\phi)) + 
 \frac{R^2}{2} \,B (f(\phi)) \,\right)  =  0. \hspace{3mm}
\end{eqnarray}
In this way, we find that
\begin{equation}
 \Lambda (x ; C_1) \ - \ \Lambda (x ; C_2) \ = \ 0.
\end{equation}
which proves the {\it path-independence} of the gauge function $\Lambda (x)$
defined by Eq.(\ref{Eq:gauge_function_Lambda}) if the closed path
does not encircle the solenoid.
This in turn implies that the line-integral connecting the two space-time points
is invariant under continuous deformation of the path connecting these 
two points as long as the curve does not cross the solenoid in the
process of deformation.
It appears to be a natural generalization
of the infinitely-long solenoid problem with time-independent magnetic flux
to more general case of time-varying magnetic flux.
Alternatively, we may be able to say that the {\it topological nature} of the
Aharonov-Bohm phase shift survives even in the case of time-varying
magnetic flux.


\end{document}